\documentclass{svproc}
\usepackage{url}

\usepackage{graphicx}
\usepackage{lineno}

\begin{document}
\mainmatter              
\title{Topological studies of charged particle production and search for jet quenching effects in small collision systems with ALICE}
\titlerunning{Topological studies of charged particle production with ALICE}  
%
\author{Sushanta Tripathy\footnote{\email{Email: sushanta.tripathy@cern.ch}\\presently at INFN - sezione di Bologna, Bologna BO, Italy}(For the ALICE collaboration)}
\authorrunning{S. Tripathy (for the ALICE collaboration)} 
%
\tocauthor{Sushanta Tripathy}
\institute{Instituto de Ciencias Nucleares, UNAM, Mexico City, Mexico}

\maketitle              

\begin{abstract}
Results for high multiplicity pp and p--Pb collisions at the LHC have revealed that these small collision systems exhibit features of collectivity. To understand the origin of these unexpected phenomena, the relative transverse activity classifier ($R_{\rm{T}}$) can be exploited as a tool to disentangle soft and hard particle production, by studying the yield of charged particles in different topological regions associated with transverse momentum trigger particles. This allows to study system size dependence of charged particle production of different origins and in particular search for jet-quenching effects. Here, results on the system size and $R_{\rm{T}}$ dependence of charged particle production in pp, p--Pb and Pb--Pb collisions at $\sqrt{s_{\rm NN}}$ = 5.02 TeV are presented.

\end{abstract}
\section{Introduction}
Recent measurements by ALICE~\cite{ALICE:2017jyt} show a smooth increase of strange to non-strange particle ratios across pp, p--Pb and Pb--Pb collisions as a function of charged-particle multiplicities at the LHC. This universal scaling with particle multiplicity may point towards a common underlying physics mechanism across collision systems. However, no onset of jet quenching effects has been observed so far the smaller pp and p--Pb systems~\cite{Nagle:2018nvi}. To disentangle the phenomena of soft (underlying event) and hard (jet induced) particle production, the relative transverse activity classifier ($R_{\rm{T}}$), an event shape observable, can be exploited as a powerful tool. Here, the production of light flavor charged hadrons for different classes of $R_{\rm{T}}$ in pp, p--Pb and Pb--Pb collisions at $\sqrt{s_{\rm NN}}$ = 5.02 $\textrm{TeV}$ are reported. Also, we present a search for jet quenching behavior in small collision systems.

\section{Relative Transverse Activity Classifier ($R_{\rm{T}}$)}
Using $R_{\rm{T}}$ proposed in~\cite{Martin:2016igp}, the final-state particle production can be studied as a function of varying underlying events. To define $R_{\rm{T}}$, the analysed events are required to have a leading trigger particle above a certain $p_{\rm T}$. Relative to the leading rigger particle, an event can be classified into three different azimuthal regions. Assuming $\phi_{\rm t}$ as the azimuthal angle for the trigger particle and $\phi_{\rm a}$ as the azimuthal angle of the associated particles, the regions can be classified as the following,
\begin{itemize}
\item Near side: $|\phi_{\rm t} - \phi_{\rm a}| < \frac{\pi}{3}$ 
\item Away side: $|\phi_{\rm t} - \phi_{\rm a}| > \frac{2\pi}{3}$
\item Transverse side: $\frac{\pi}{3} \leq |\phi_{\rm t} - \phi_{\rm a}| \leq \frac{2\pi}{3}$
\end{itemize}
The near side is dominated by the jet activity related to the trigger particle. Since jets are typically produced as pairs back-to-back in azimuthal angle, the away side will contain some of these back-scattered jets. The transverse side is dominated by particle production in the underlying events (UE). Both the near and away side also contain similar UE production as in transverse side. Thus, one can subtract the UE from near and away side by subtracting the yield in transverse side, see Section~\ref{res}. The leading-$p_{\rm T}$ selection of  8 $< p_{\rm T}^{\rm trig.} <$ 15 GeV/$c$ ensures that the number density in the transverse region remains almost independent of leading particle $p_{\rm T}$~\cite{Acharya:2019nqn} and reduces the impact of possible elliptic flow on the measurements. $R_{\rm T}$ is defined as~\cite{Martin:2016igp,Ortiz:2017jaz},
\begin{eqnarray}
R_{\rm T} = \frac{N_{\rm ch}^{\rm TS}}{\langle N_{\rm ch}^{\rm TS} \rangle},
\label{eq2}
\end{eqnarray}
where, $N_{\rm ch}^{\rm TS}$ is the charged particle multiplicity in the transverse side. The events with $R_{\rm T} \rightarrow$ 0 are the events expected to be dominated by jet fragmentation.

\section{Results and Discussion}
\label{res}

\begin{figure}[ht!]
\centering
\includegraphics[width=29pc]{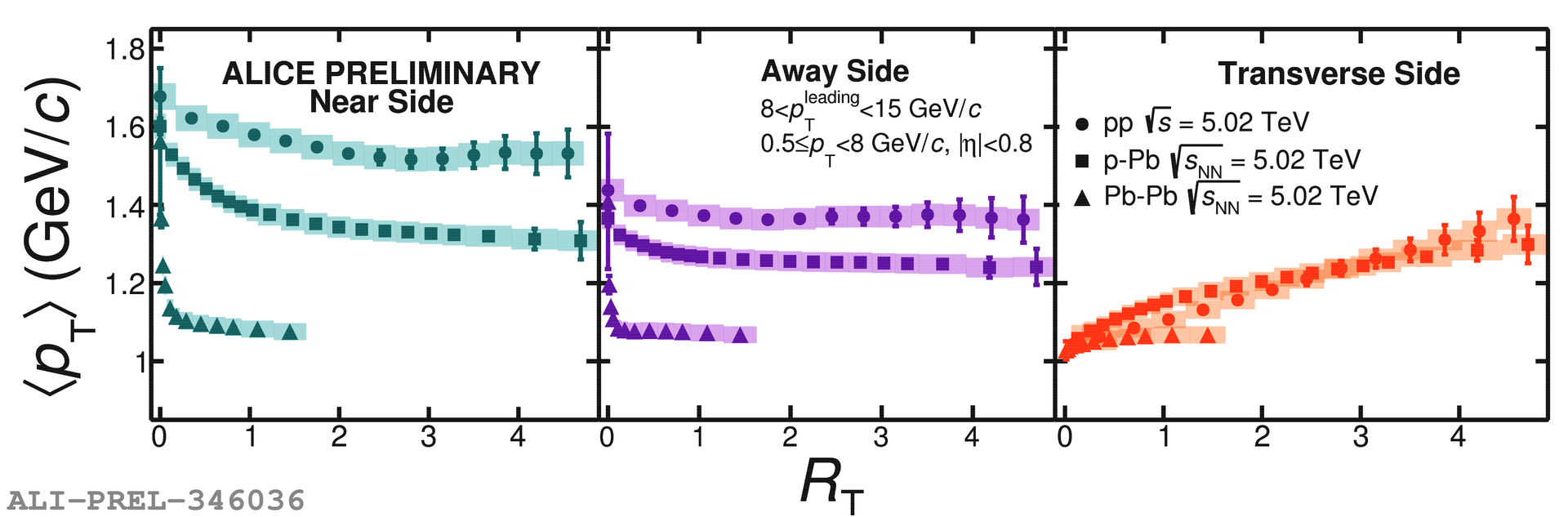}
\caption{\label{fig1} System size dependence of $\langle p_{\rm T}\rangle$ for charged-particles as a function of $R_{\rm T}$ in the near (left), away (middle), and transverse (right) sides.}
\end{figure}

Figure~\ref{fig1} shows $\langle p_{\rm T}\rangle$ of charged-particles as a function of $R_{\rm T}$ in the near (left), away (middle), and transverse (right) sides for pp, p--Pb and Pb--Pb collisions. The measurement of charged-particles follows a similar procedure as described in Ref.~\cite{Acharya:2018qsh}. The near and away side $\langle p_{\rm T}\rangle$ for pp and p--Pb collisions decreases at low-$R_{\rm T}$ and it saturates for high-$R_{\rm T}$. This behavior indicates that the contribution from the near and away side jet dominates at low-$R_{\rm T}$ and the soft particle production starts contributing in high-$R_{\rm T}$ region. Another interesting observation to note that the values of $\langle p_{\rm T}\rangle$ are similar for all systems for $R_{\rm T}\rightarrow$ 0. One would naively expect this behavior as this region has very little contribution from soft particles i.e. UE. The $\langle p_{\rm T}\rangle$ for transverse side increases with $R_{\rm T}$ as the UE increases with increase in $R_{\rm T}$. For large $R_{\rm T}$, the $\langle p_{\rm T}\rangle$ approaches to a similar value in all three topological regions for a given system as they are mostly dominated by the UE.

\begin{figure}[ht!]
\centering
\includegraphics[width=14pc]{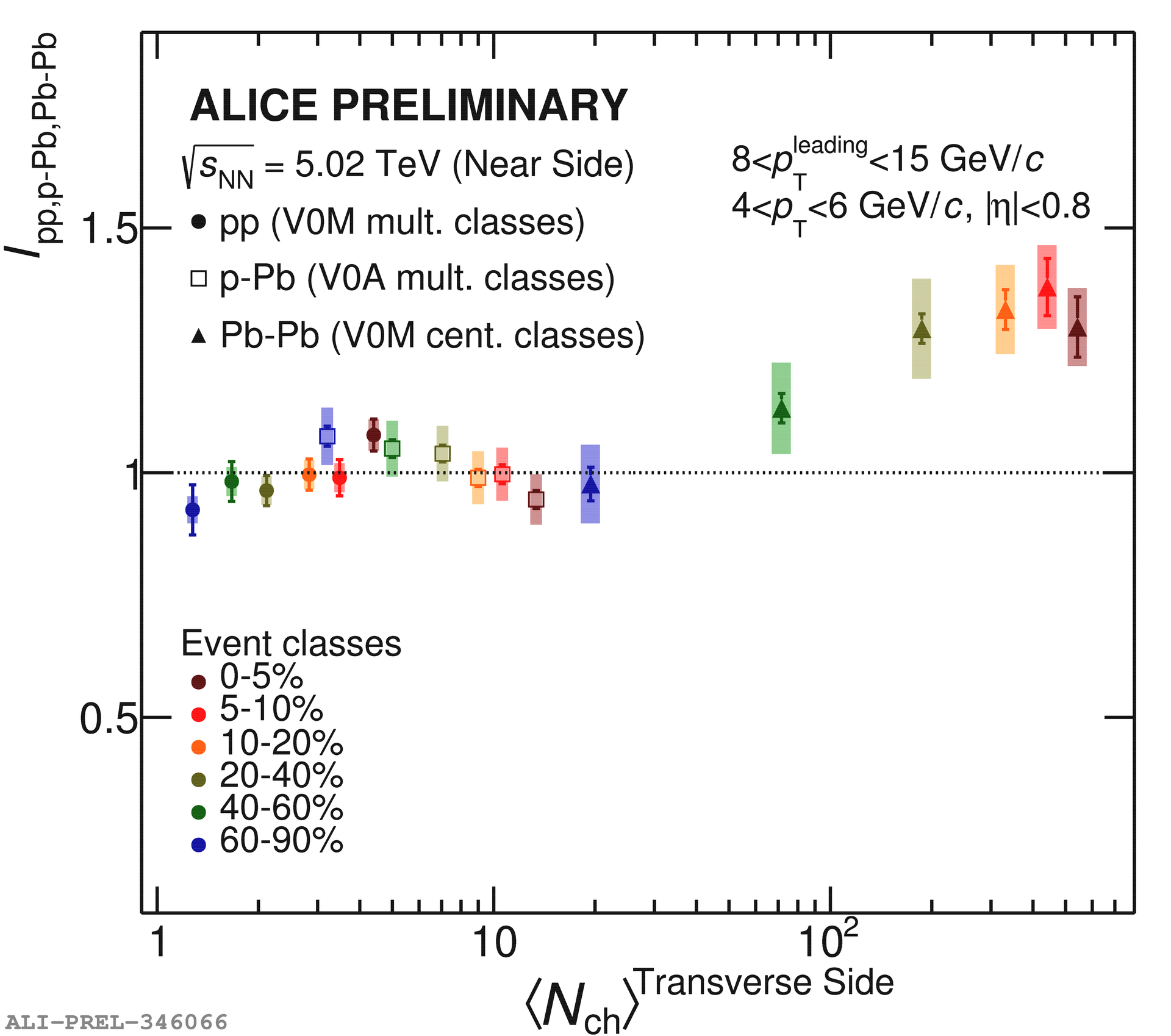}
\includegraphics[width=14pc]{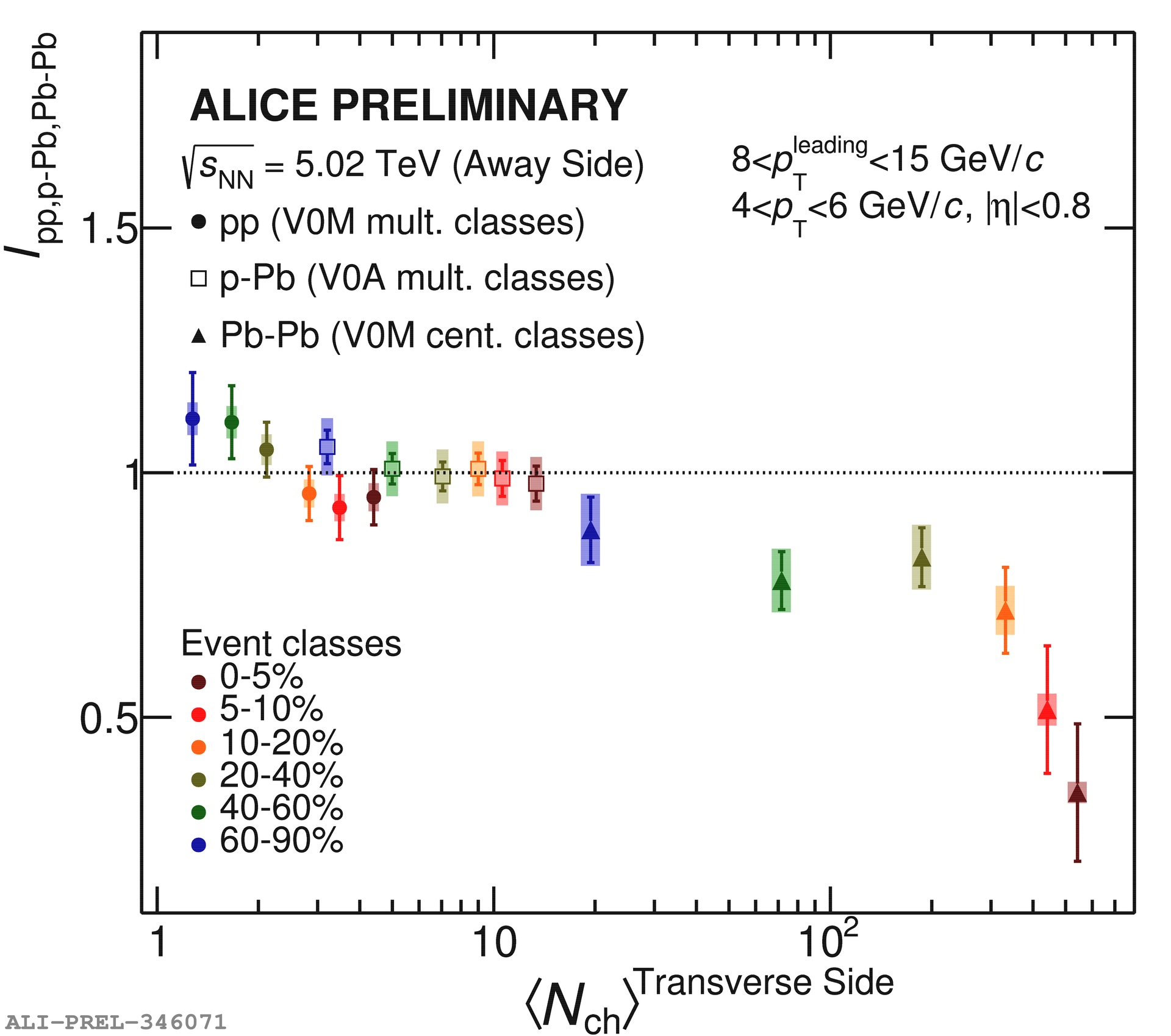}
\caption{\label{fig2} $I_{\rm pp,p-Pb,Pb-Pb}$ as a function of $\langle N_{\rm ch}^{\rm TS} \rangle$ in different V0M/V0A multiplicity classes for the near (left) and away (right) side in pp, p--Pb, and Pb--Pb collisions at $\sqrt{s_{\rm NN}}$ = 5.02 TeV.}
\end{figure}

To investigate the presence of jet-quenching effects in small collision systems, $I_{\rm pp,p-Pb,Pb-Pb}$, an observable which is calculated from the yields of different topological regions, is calculated as a function of $\langle N_{\rm ch}^{\rm TS} \rangle$ for different V0M (V0A) multiplicity classes of pp and Pb--Pb (p--Pb) collisions. The $I_{\rm pp,p-Pb,Pb-Pb}$ is a similar quantity ($I_{\rm AA}$) calculated as in Ref.~\cite{Aamodt:2011vg}. The $I_{\rm pp,p-Pb,Pb-Pb}$ is expected to be highly sensitive to medium effects. A suppression of this obervable in the away side would indicate the presence of jet quenching, while an enhancement in the near side would indicate a bias due to trigger particle selection and/or presence of medium effects.  It is defined as the ratio of yield in the near or away region (after subtraction of yield in transverse side) in different collision systems to the yield in the near or away region in minimum bias pp collisions. It can be expressed as,

\begin{equation}
I_{\rm pp,p-Pb,Pb-Pb} = \frac{Y^{\rm pp,p-Pb,Pb-Pb} - Y^{\rm pp,p-Pb,Pb-Pb}_{\rm TS}}{Y^{\rm pp~min. bias} - Y^{\rm pp~min. bias}_{\rm TS}}.
\end{equation}

Here, $Y$ represents the integrated yield of charged particles in a particular topological region. For these results we have not made a direct selection on $N_{\rm ch}^{\rm TS}$, as the direct selection on $N_{\rm ch}^{\rm TS}$ biases the near and away side yields~\cite{Ortiz:2020dph}. Thus, the events are selected based on the forward rapidity estimator (V0M for pp and Pb--Pb collisions and V0A for p--Pb collisions) and the corresponding $N_{\rm ch}^{\rm TS}$ are calculated for each multiplicity class. Figure~\ref{fig2} shows the $I_{\rm pp,p-Pb,Pb-Pb}$ in the range 4 $< p_{\rm T}^{a} < $ 6 GeV/$c$ as a function of $\langle N_{\rm ch}^{\rm TS} \rangle$ in different V0M/V0A multiplicity classes for the near (left) and away (right) side in pp, p--Pb, and Pb--Pb collisions at $\sqrt{s_{\rm NN}}$ = 5.02 TeV. Here, $p_{\rm T}^{a}$ is the transverse momentum of associated particles with respect to a leading trigger particle.  The values of $I_{\rm PbPb}$ for most central and most peripheral Pb--Pb collisions show similar trends as reported by ALICE in Ref.~\cite{Aamodt:2011vg} at Pb--Pb collisions at $\sqrt{s_{\rm NN}}$ = 2.76 TeV. In small collision systems, no enhancement (suppression) of $I_{\rm pp,p-Pb}$ is observed in near (away) sides for pp or p--Pb collisions within uncertainties. This indicates the absence of jet-quenching effects in small collision systems or if any, the jet-quenching effects are very small to be detected in the measured $\langle N_{\rm ch}^{\rm TS} \rangle$ ranges.

\section{Summary}

In summary, using $R_{\rm T}$ one can vary the magnitude of the underlying event contribution and study the final state particle production in different topological regions. The system size dependence of charged-particle production indicates that the contribution from the near and away side jet dominates at low-$R_{\rm T}$. For high-$R_{\rm T}$, the $\langle p_{\rm T}\rangle$ approaches a similar value in all three topological regions for a given collision system. In contrast to Pb--Pb collisions, no suppression of $I_{\rm pp,p-Pb}$ is observed in the away side for pp and p--Pb collisions, which indicates the absence of jet-quenching effects for small collision systems or if any, the jet-quenching effects are very small to be detected in the measured  $\langle N_{\rm ch}^{\rm TS} \rangle$ ranges.

\section*{Acknowledgements}
S.T. acknowledges the support from CONACyT under the Grant No.A1-S-22917 and postdoctoral fellowship of DGAPA UNAM.

%
%

\end{document}